\documentstyle[eqsecnum,aps]{revtex}
\begin{document}
%\twocolumn[\hsize\textwidth\columnwidth\hsize\csname@twocolumnfalse\endcsname

\title
{Correspondence between time-evolution dynamics of a tumor and an
attractively interacting Bose-Einstein Condensate with feeding and 
dissipation.}

\author{P. K. Biswas$^{*,\#}$ and M. T. T. Pacheco$^*$}
\address{ 
{}$^*$Divis\~ao de Engenharia Biom\'edica,\\ Instituto de 
Pesquisa e Desenvolvimento, Universidade do Vale do Paraiba;\\
S\~ao Jos\'e dos Campos 12244-000, SP, Brasil\\
{}$^\#$Departamento de F\'{\i}sica, Instituto Tecnol\'ogico de 
Aeron\'autica, CTA\\
S\~ao Jos\'e dos Campos 12228-901, SP, Brasil\\
email: biswas@fis.ita.br/biswas@univap.br\\}

%\date{\today}
\maketitle

%\begin{abstract}
\section{summarry}
The morphology and time-evolution dynamics of tumors are expected to depend 
heavily on
the detailed balance of the overall physics of the cell assembly
(e.g., the kinetic pressure, the cell-cell interaction, and the external
trapping by the tissue) and the biological processes of mitosis, 
necrosis, etc. Here, for the first time, we include such a detailed balance
in a theoretical model for tumor by exploiting an {\it ab initio} 
mathematical framework of the atomic Bose-Einstein Condensate (BEC)
with feeding and dissipation. 
We show that the Gross-Pitaevskii equation, which describes the
many-body atomic BEC characteristics,
indeed explains the detailed features of a prevascular tumor culture data
with a characteristic length scaling.
The agreement suggests the prevascular carcinoma may be a natural analog
to BEC and predicts an intercellular wave connecting the cells.
%\end{abstract}

\section {Introduction} 
Studies on tumors reveal that they are finite-sized and 
inhomogeneous assembly of cancer cells which appear in a bound spheroid
form for prevascular carcinomas \cite{sutherland,dalen,folkman}. 
We emphasize that,
whether in vivo or in vitro they certainly do experience an 
external trapping 
potential exerted by the surrounding medium (suspension or tissue).
So, at any instant of time, the tumor morphology might be an outcome of 
a balance of the overall physics of the cell assembly: the kinetic pressure,
the intercellular 
interaction, and the external trapping potential. Consequently, the tumor
evolution is expected to be guided by the balance of these physical forces
together with the biological processes: 1) cell proliferation or mitosis, 
2) impediment to mitosis (pre-vascular case) in the 
region of higher cell density, 3) and consequent appearance of quiescent 
cells and cell death due to nutrient level lower than that needed to 
proliferate or sustain life, etc.

Theoretical studies, however, mainly employ  first-order
differential equations to simulate the number or volume growth
by shooting various mitosis and necrosis rates and 
immunological reactions \cite{kuznetsov,fodiff},  
spatio-temporal equations (differential and integro-differential)
\cite{tracqui,burges,adam} to take into account
the diffusion of nutrients, and cellular
automation (lattice simulation) models \cite{stott,dormann,mombach}
to mimic the growth pattern considering empirical cell-cell interactions. 
None of the previous attempts emphasizes this balance of
the physical potentials and the biological reactions and it remains a 
challenge to embed them together in a non-empirical framework.

Looking at the recently discovered \cite{anderson,davis,bradley} 
new form of matter, the Bose-Einstein
Condensate (BEC), and their time-evolution dynamics for an attractive 
atom-atom interaction \cite{kagan,victo,bistobi1}, 
we understand that the underlying physical
factors (kinetic pressure, atom-atom interaction, external trapping) of the 
trapped atomic assembly are quite similar to those expected to be
present in the tumors. Also, atomic BEC states are  
finite-sized and inhomogeneous, like the tumors. The feeding to the 
condensate and dissipation from it can mimic the role of mitosis and 
necrosis. Consequently, we conjecture
that the Gross-Pitaevskii equation which successfully
describes the BEC matter configuration and their time-evolution, could,
in some form, be employed to the tumor of cells.

Indeed, we show here that a scaled form of the Gross-Pitaevskii equation,
can explain all the intricate features of the in-vitro tumor culture 
data \cite{folkman}.

\section{The Gross-Pitaevskii (GP) Equation}
The nonlinear GP equation is regarded as a classical field approximation
to the many-body Heisenberg equation \cite{dalfovo,huang}. 
Also, it
can be derived from a field Lagrangian \cite{jpb} with self-interaction
terms. In a non-conservative form it is represented as \cite{kagan,victo}:
\begin{eqnarray}\label{gp1}
i\hbar\frac{\partial}{\partial t}\Phi({\bf r},t)=
\biggr[-\frac{\hbar^2}{2m}\nabla^2
+\frac{4\pi\hbar^2a}{m}|\Phi|^2
+V_{trap}
+ V_{nc}\biggr]\Phi({\bf r},t)
\end{eqnarray}
where $\Phi$ is a classical field approximation to the Heisenberg field
operator $\hat{\Psi}$ and 
the atom-atom interaction is considered in a mean-field approximation
\cite{dalfovo}
\begin{eqnarray}\label{g}
V({\bf r'-r})=\frac{4\pi\hbar^2a}{m}\delta({\bf r'-r})
\end{eqnarray}
$m$ is the mass of an atom; $a$
is the $s$-wave scattering length also known as the hard-sphere radius
\cite{huang}
 of the interaction potential; and $h$ is the Planck's constant 
($\hbar=h/2\pi$). The normalization of the order parameter $\Phi$ provides
the number of atoms $N(t)$ at any time $t$: 
$N(t)=\int|\Phi(r,t)|^2d^3r$. $\nabla^2$ is the Laplacian operator;
$V_{trap}$ is the external trap potential which can be of any form:
harmonic, cylindrical, square well, etc.
$V_{nc}$ is a non-conservative term
in the Hamiltonian which has been introduced by Kagan et al \cite{kagan}
as
\begin{eqnarray}\label{nc1}
V_{nc}=i\frac{\gamma}{2}\hbar\omega-2i\xi\biggr(\frac{4\pi\hbar a}
{m\omega}\biggr)^2\hbar\omega|\Phi|^4
\end{eqnarray}
for a harmonic trap: $V_{trap}=\frac{1}{2}m\omega^2 r^2$,
where $\omega$ is the frequency of the trap.  
The first term of (\ref{nc1}) corresponds to growth in the condensate by 
feeding from
thermal cloud, and the second term corresponds to loss of atoms due to
atomic dimer formation from three-body interactions. $\gamma$ and $\xi$
are parameters.

Defining an oscillator length 
$a_{ho}=\sqrt{\hbar/(2m\omega)}$ (normally it is defined as 
$a_{ho}=\sqrt{\hbar/(m\omega)}$), and a dimensionless length, time and order
parameter as
$x=r/a_{ho}$, $\tau=\omega t$, $\phi(x,\tau)=\sqrt{8\pi|a|}\Phi r$, one
represents eqn.\ref{gp1} in the following dimensionless form:
\begin{eqnarray}\label{gpnd}
i\frac{d\phi}{d\tau}=\biggr[-\frac{d^2}{dx^2}+\frac{1}{4}x^2
-\frac{|\phi|^2}{x^2}-2i\xi\frac{|\phi|^4}{x^4}+i\frac{\gamma}{2}\biggr]
\phi
\end{eqnarray}
In terms of $\phi$,
the number of atoms and the 
mean-square radius are given by: 
\begin{eqnarray}\label{nbec}
N(\tau)&=&
\frac{4\pi}{8\pi|a|}a_{ho}\times
\int|\phi(x,\tau)|^2dx
=\frac{1}{2|a|}a_{ho}\times n(\tau) \\
\langle R^2(\tau)\rangle &=& 
a_{ho}^2 \times
\frac{\int x^2|\phi(x,\tau)|^2 dx}{\int |\phi(x,\tau)|^2 dx}=a_{ho}^2 
\langle x^2(\tau) \rangle\label{rbec}.
\end{eqnarray}
where $n(\tau)=\int|\phi(x,\tau)|^2 dx$.
The time-independent conservative form of eqn.\ref{gpnd} 
is given by \cite{burnett}:
\begin{eqnarray}\label{gpti}
\biggr[-\frac{d^2}{dx^2}+\frac{1}{4}x^2
-\frac{|\varphi|^2}{x^2}\biggr]\varphi
=\beta\varphi
\end{eqnarray}
where $\phi$ in eqn.\ref{gpnd} is taken as $\phi(x,\tau)\equiv 
\mbox{exp}(-i\beta \tau)
\varphi(x)$;
with $\beta=\mu/\hbar\omega$; $\mu$ representing the chemical 
potential or average single-particle energy. 
This equation (\ref{gpti}) has various stable, metastable, or unstable 
solutions \cite{jpb}
corresponding to various values of $\beta$. 
Each solution corresponds to a particular value of $N|a|/a_{ho}$
(see eqn.\ref{nbec}).
The time-evolution of the condensate is studied by feeding the solution
of this equation as an input to eqn.\ref{gpnd}.

\subsection{Compatibility of the GP equation}
The GP eqn.\ref{gp1}, which is obtained by replacing the field operator
$\hat{\Psi}$ by a classical order parameter $\Phi$ in the many-body 
Heisenberg equation \cite{bogoliov},  
constraints the theory to remain valid till the condensate population
is very high so that the annihilation and the creation operators can be 
considered as $c$-numbers. 
In atomic BEC this condition is achieved in the limit of 
$T\rightarrow 0$ when most of the particles are expected to occupy the 
lowest energy state making its population very high. 
A tumor, which we call a 'cell-condensate', is represented by a density
of about $\sim 10^6-10^8$ cells per cc. and the size varies from fraction
of a few mm to a few cm at a temperature of $37-39^oF$.

  The MFA is valid when the condensate is "sufficiently dilute" so that
$\rho |a|^3<<1$  ($\rho$ is the density; $a$ is the $s-$wave scattering
length). 
For cells in a temperature of $37-39^oF$, it is not  expected that the
cell-cell interaction be fully represented by the $s$-wave scattering 
length alone. Also, the latter is not known for the cells. However, to deal
with the huge number of complex cancer cells inside the tumor, the mean 
field approximation appears to the most suitable choice at the moment, 
if the condition $\rho |a|^3<<1$ does not appear nonconducive, where $|a|$
is properly simulated to represent the cell-cell interaction.
We approximately fix it by considering that the hard-sphere potential range
is of the same order of the 
effective cell radius and take $|a|\approx \bar r$ as the interactions are
of short range. Then from the measured \cite{folkman} total number of cells
$N(t)$ for the V-79 spheroids and its average 
volume $V(t)$ at an instant $t$,  we see that 
$\rho |a|^3 \sim 10^{-2}$
except at the very beginning (day-10) where it is still less than $1$,
but marginally.

\subsection{Scaling of the GP equation}
The essential two parameters of the dimension of length,
which prevail in a BEC matter, are 1) $\lambda-$the thermal wavelength 
($=h/\sqrt{2\pi m k T}$;
$k=$Boltzmann constant and $T=$condensate temperature) 
and 2) $v^{1/3}-$the average interparticle separation.
For an ideal case, $\lambda$ satisfies the relation \cite{huang}: 
\begin{eqnarray}
\frac{\lambda^3}{v}=g_{3/2}(1)+\frac{\lambda^3}{V}\frac{z}{1-z}
\ge g_{3/2}(1).
\end{eqnarray}
where $z$ is the fugacity of the atoms, $V$ is the volume of the
container, and $g_{3/2}(1)=2.612$.
In a tumor, the parameter corresponding to the interatomic separation 
$v^{1/3}$ in a BEC matter is the intercellular separation ${\bar v}^{1/3}$.
Any correspondence of a tumor of cells to a BEC of atoms should then be 
reflected through a length scaling of the form $({\bar v}/v)^{1/3}$.
From eqs.\ref{nbec} and \ref{rbec}, we see that 
the number of atoms and the condensate volume are defined by the
dimensionless order parameter $\phi(x,\tau)$, a characteristic length 
$a_{ho}$, and the $s-$wave scattering length $a$. So, we embed the length
scaling to the characteristic oscillator length $a_{ho}$ of the GP equation
and set
\begin{eqnarray}\label{scale}
{\bar a}_{ho}=({\bar v}/v)^{1/3} a_{ho}=({\bar d}/d) a_{ho}.
\end{eqnarray}
where $d$ ($\bar d$) is the effective size of an atom (cell) in a particular
BEC (tumor) configuration. 
Consequently, we define the number of cells and the size of the tumors
by replacing $a_{ho}$ with ${\bar a}_{ho}$ in eqs.\ref{nbec} and \ref{rbec}.
\begin{eqnarray}\label{nbect}
N(\tau)&=& \frac{1}{2|{ a}|}{\bar a}_{ho}\times n(\tau)\approx
\biggr(\frac{a_{ho}}{d}\biggr)\times n(\tau) \\
R(\tau) &=& {\bar a}_{ho}\sqrt{\langle x^2(\tau)\rangle}=
\biggr(\frac{a_{ho}}{d}\biggr)\times {\bar d} \times 
\sqrt{\langle x^2(\tau)\rangle}
\label{rbect}
\end{eqnarray}
The value of $\bar d$ is obtained from the measured
data of ref.\cite{folkman} while the value of $a_{ho}/d$ is fixed by
selecting an appropriate solution of the GP equation for the short-range,
attractively interacting, and externally trapped ${}^7$Li condensate 
\cite{kagan,victo,bistobi1}.

Before presenting the results, we comment on the effect of the length
 scaling on $\lambda$, the thermal wave. Applying the scaling we see
that the intercellular separation in a tumor could satisfy a relation of
the form:
\begin{eqnarray}\label{289}
{\bar{\lambda}}^3 \ge {\bar v} g_{3/2}(1)
\end{eqnarray}
Although we do not investigate the physical existence of such an wave
length in living bodies, we mention here that any existence of such an
wave, connecting the cells
would be very useful to directly account for the observed 
ultraweak and coherent biological light emanating from tissues 
\cite{bioptn1}
and support the school of opinion that it comes from an organized energy
field which communicates within the whole organism: {\it an intercellular
communication} \cite{bioptn3}.

\section{Results and Discussions}
To demonstrate the basic features of the theory we intend to reproduce
the tumor culture
data on the viable cell count and volume on the V-79 Chinese Hamster lung
cells as measured by Folkman and 
Hochberg (FH) \cite{folkman}.

To simulate the experimental condition of the tumor spheroids 
suspended (trapped) in soft agar, we consider a agar trap defined by 
$V_{ext}=V_x\hbar\omega$, with the boundary condition $V_x=\infty$ at the 
container walls. (For a trap independent of $\omega$, the latter is a
simple parameter with the dimension of time-inverse.) 
We then solve eqn.\ref{gpti} for a bound
ground state solution for a few cells (about five) and next
evolute the tumor state in time using eqn.\ref{gpnd}.
FH has tried to maintained a constant feeding by replacing
the agar at regular intervals of 2-3 days. Similarly, we consider a
constant feeding in the Hamiltonian given by $\frac{i}{2}\gamma\hbar\omega$,
replicating a fixed mitosis rate. ($\gamma$ is a parameter representing the
strength of the feeding.) In practice, although feeding remains constant
for a prevascular tumor, the mitosis rate is slowed down with increasing
size and density of the tumor. It is now known that the
1) diffusion of nutrient become less in the central region;
2) consequently, some cells therein enter a quiescent phase where
they remain alive but non-proliferating; and
3) some of the quiescent cells die due to  nutrient 
diffusion lower than that needed for survival.
All the above mentioned negative aspects of the growth are expected to depend
on the density and the kinetic profile of the cell assembly.
We find that a non-linear density dependence proportional to 
$\rho^2(x,t)$ $(\equiv |\Phi(x,t)|^4)$ in the Hamiltonian, explains the 
intricate 
features of the viable cell count data of FH \cite{folkman} for the 
V-79 cells.

In figure-1a we present the number of cells $N(t)$ versus the time 
$t=\tau/\omega$.
First, to fix the time scale ($t=\tau/\omega$) we observe
that $\omega$ corresponds to a value $ \omega \sim 1$ per day and thus we
set $\omega=1/(24\times 60\times 60) sec^{-1}$. 
From figure-1a
we find that 1) initially the number increases with a steep exponential 
fashion; 2) it attains a local
peak at around day-26 (solid line); 3) here it suffers a partial collapse;
4) after completing the collapse, it again starts a strong exponential 
growth at around day-35 but for a short 
duration and gradually it gets stabilized with a viable count around
$N\sim 10^5$. From figure-1b (reproduced from figure 3b of 
ref.\cite{folkman}) we see that 
all the characteristic features
described above are present in the measured data of FH including the
(approximate) positions of the peak and collapse.

Among the crucial features in the measured data (figure 1b), 
{\bf we highlight the partial
collapse after day 24 and the short but strong regrowth without an 
angiogenesis near day-30}. This is a typical characteristic of an atomic
BEC with
atom-atom short-range nonlinear attractive interaction and three-body 
recombination
losses \cite{kagan,victo,bistobi1}.
Studying the density profile (see figure 2), we find that with 
growing number, the central density increases the most.
The tumor appears to form a multilayer system as can be seen from
figure 2c. The increasing central zone density reaches a maximum
around day 24 when it triggers a contraction or partial collapse of
the system, 
exactly similar to that reported for the many-body atomic 
condensate \cite{kagan}. 
In the process of 
contraction there could  be a substantial loss of cells in that 
region as can be seen from the density profile at day-28 in figure 2d.
This loss in turn would diminish the 
attractive pseudopotential resulting a fresh rapid expansion of the system 
which is evident both in figure 1a and and figure 1b around day-30
(for a similar event in an atomic BEC see around $\tau\sim 5.0$ in figure-1
of ref.\cite{kagan}). The inviable and dead cells in the 
central region are referred as a necrotic core \cite{folkman,stott}.
The fluctuation in number in the steady (dormant) state is a 
repeatation of the expansion and the contraction process mentioned above. 
In the following we discuss the volume evolution of the V-79 spheroids which
represents some apparent anomaly with their number evolution.

Comparing the measured data on the viable and total cell counts in figure
1b and the spheroid volume (circles with
error bar in figure 3), we note that 
1) while both the cell numbers 
approximately stabilize 
since day 50-55, the diameter of the tumor at that time is about one-third
of its stabilized value of $3.6\pm0.5$mm (figure 3). 
2) From day-55 
onwards, the size triples against a constant (approx.) number.
From the V-79 data, we evaluate the effective cell size
and find that it varies from ${\bar d}\approx 65\mu$m near day 10, to 
${\bar d}\approx 10\mu$m near
day 42, to finally get stable with a value of ${\bar d}\approx 40\mu$m at
around day 200.
The abrupt decrease in the effective cell size from 
$65\mu$m to $10\mu$m might be due to the fact that in the initial phase
of aggressive mitosis (day 0-24), the cells might be undergoing 
mitosis before they could double their individual size. 
The volume gain in the latter phase (day 50-160) against a approximately
fixed number of cells may be attributed to a gain in the effective and
individual cell size as we cannot expect any other mechanism of volume 
expansion from a lower attractive pseudo potential as the number is fixed.

To check this assertion, 
we approximately formulate the variation in the average cell size as:
\begin{eqnarray}
{\bar d}(t)= d_m^-(t) + d_g^+(t)
\end{eqnarray}
where
$d_m^-$ ($=d_0e^{-c_1 t}$), is a rapidly diminishing variable of size
representing aggressive mitosis and $d_g^+$ ($=d_f(1-e^{-c_2 t^{c_3}})$),
is a rather slowly increasing variable of size. Where $d_0=65\mu$m,
$d_f=40\mu$m and we choose the value of the parameters
$c_1,c_2,c_3$ such that we obtain a minimum average size
$\sim 10\mu$m and a stabilization size of $40\mu$m (after day 160).

The variation of $\bar d$ will have no 
effect in eqn.\ref{nbect} as it is expected to
cancel from a similar variation in $|a|$ as can be seen from our
consideration $|a|\approx \bar r$ ($=\bar d/2$). However, it will affect
the volume as
we see from eqn.\ref{rbect}. In our calculation $\phi$ represents 
the `wave function' for the proliferating (viable) cells and so
$\sqrt{\langle x^2 (\tau)\rangle}$ in eqn.\ref{rbect}, will provide the 
average spread of the proliferating cells which 
occupy the outermost boundaries (the necrotic mass is concentrated
in the center). Consequently, eqn.\ref{rbect} is expected to give the
size of the tumor provided the effective cell size $\bar d$ is appropriate.
Theoretical results are shown
in the solid curve in figure 3 and the experimental data of 
ref\cite{folkman} are shown as circles with the error bars. 
The time-variation in the tumor size agree quite well with the measured 
data vindicating the possibility of the above explanation to the 
volume gain against a fixed number count. The crucial aspect is the agreement
in the number count of proliferating (viable) cells and the stabilized tumor
size.

Present investigation reveals a correspondence between a many-body
atomic system in a condensed state to a multicellular tumor spheroid.
Invoking a dimensionless form of the GP equation, and 
considering the tumors as bound and trapped assembly of cancer cells 
with nonlinear intercellular interactions, we explain
the long-standing time-evolution data \cite{folkman} on spherical carcinoma
with all their essential features. 
This raises the 
possibility of viewing a prevascular carcinoma as a natural analog to BEC. 
The response of the avascular tumors to the GP equation and to the 
nonlinear
physics of BEC matters mark a significant new development of the 
cell-proliferation dynamics and is expected to usher a new
era in the future studies of tumor evolution. 
The involved scaling of the BEC matter lead to a matter wave
with wavelength of the order of intercellular separation. The physical
presence of such an wave will provide a foundation to the biophoton
hypothesis and establish an intercellular communication, conjectured 
previously \cite{bioptn3}.

\section{Acknowledgements}
Authors acknowledge the partial financial support of FAPESP. PKB 
thanks J. S. E. Germano
for scanning and reproducing  figure 1b and for his interest in the 
work. PKB visioned the model and acknowledges valuable discussions
with T. Frederico on nonlinear BEC and with W. Ribeiro, N. S. Silva,
J. C. Cogo, C. P. Soares, C. Chavantes, 
Socrates, and F. A. S. Carvalho about various biological aspects of the
tumors and thanks library staff A. M. Carvalho for urgently arranging
some reference articles from outside.\\

Authors do not have any competing financial interest.\\

Correspondence and requests for material should be addressed to P.K.B
(email: biswas@fis.ita.br)\\

{\Large Figure captions}:

{\bf Figure 1a:} Number evolution for the proliferating (viable) V-79 cells
in a spherical carcinoma in agar. 

{\bf Figure 1b:} Figure reproduced from Folkman and Hochberg \cite{folkman}
by scanning. V-79 data: Solid curve- viable cell count; dashed curve- total
cell count.

{\bf Figure 2:} Density ($|\phi(x,\tau)/x|^2$) distribution of the V-79 
spherical carcinoma in their early stages. a) initial tumor with five
cells at $\tau=0$, b) $\tau=10$,
c) $\tau=24$, and d) $\tau=28$.

{\bf Figure 3:} Volume evolution of V-79 spherical carcinoma. Circles with
error bars are the measured data reproduced from figure 3b of 
ref.\cite{folkman}; Solid curve - present results.

\end{document}